\documentclass{article}
\usepackage{frascatiphys,here,graphicx,subfigure}

\usepackage{relsize}
\def\bbrExp{\mbox{\slshape B\kern-0.1em{\smaller A}\kern-0.1em
    B\kern-0.1em{\smaller A\kern-0.2em R}}}
\begin{document}
\title{
CHARM SPECTROSCOPY AT \bbrExp}
\author{
Vincent Poireau\\
{\em LAPP, IN2P3/CNRS et Universit\'e de Savoie, Annecy-Le-Vieux, France} \\
}
\maketitle
\baselineskip=11.6pt
\begin{abstract}
We present a mini-review on charm spectroscopy at the \bbrExp\ experiment. We first
report on the $c\bar{s}$ meson spectrum, and present precise measurements of the
$D_{s1}(2536)$ meson as well as the properties of the many new states discovered since
2003 ($D_{s0}^*(2317)$, $D_{s1}(2460)$, $D_{sJ}^*(2860)$, and $D_{sJ}(2700)$ mesons). We
then discuss about charmed baryons observed recently in the \bbrExp\ experiment:
$\Omega_c^0$ and $\Omega_c^{*0}$ $css$ baryons, $\Lambda_c(2940)^+$ $udc$ baryon and the
$\Xi_c$ $usc/dsc$ baryons.
\end{abstract}
\baselineskip=14pt

\section{Introduction}

Observations of a long list of new meson and baryon resonances have been recently
reported by the \bbrExp, Belle and CLEO experiments. We present here the new resonances
observed in the $c\bar{s}$ and charmed baryon sectors by the \bbrExp\ experiment.

In this short review, we do not present results on $c \bar{u}$ and $c \bar{d}$ resonances
($D$, $D^*$, $D^{**}$ mesons) and on $c \bar{c}$ states (charmonium or charmonium-like
state).

Analyzes presented here were performed using data collected at the $\Upsilon(4S)$
resonance with the \bbrExp\ detector\cite{ref:babar}, located at the PEP-II asymmetric
energy $e^+e^-$ collider. From 1999 to 2007, the \bbrExp\ experiment recorded a
luminosity of 432~fb$^{-1}$ at the $\Upsilon(4S)$ peak and 45~fb$^{-1}$ at 40~MeV below
the peak. This corresponds to about $475.10^6$ $B \bar{B}$ pairs and $620.10^6$ $c
\bar{c}$ pairs. The analyzes presented here use only a subset of the total recorded
luminosity.

\section{$c\bar{s}$ mesons}

Before 2003, only four $c\bar{s}$ mesons were known: two S-wave mesons, $D_s$ ($J^P =
0^-)$ and $D_s^*~(1^-)$, and two P-wave mesons, $D_{s1}(2536)~(1^+)$ and
$D_{s2}(2573)~(2^+)$. The masses predicted by the potential model\cite{ref:potential}
were in good agreement with the measured masses. The potential model predicted also two
other broad states (width of a few hundred of MeV) at masses between 2.4 and
2.6~GeV/$c^2$.

\subsection{$D_{s1}(2536)$ meson}

With the discovery of additional $c\bar{s}$ mesons, a comprehensive knowledge of all
known $D_s$ mesons is mandatory. As of 2006, the properties of the $D_{s1}(2536)$ meson
were not perfectly known: the width was determined to be less than 2.3~MeV at 90\%
confidence level, and the quantum numbers were only inferred.

In Ref.\cite{ref:Ds1}, a high precision measurement of the mass and of the width was
performed, using events in the $c \bar{c}$ continuum. Reconstructing inclusively the
decay $D_{s1}(2536) \to D^{*+} K^0_{S}$, with $D^{*+} \to D^0 \pi^+$ and $D^0$ decaying
either to $K^- \pi^+$ or to $K^- \pi^+ \pi^- \pi^+$, one obtain a mass of $(2534.85 \pm
0.02 \pm 0.40)$~MeV/$c^2$ (where the first error is statistical and the second is
systematic). The error on the mass is dominated by the uncertainty on the $D^{*+}$ mass.
Furthermore, the width was measured at a value of $(1.03 \pm 0.05 \pm 0.12)$~MeV. This is
the first time that a direct measurement of the width is given, rather than just an upper
limit.

Additionally, in another analysis\cite{ref:DDK}, the $D_{s1}(2536)$ meson was
reconstructed exclusively in $B$ decays, with $B \to D^{(*)} D_{s1}(2536)$ (8 modes in
total) followed by $D_{s1}(2536) \to D^* K$. A total of $182 \pm 19$ events is seen,
which gives an observation at the 12$\sigma$ significance level (where $\sigma$ is the
68\% C.L. standard deviation). With this method, a mass of $(2534.78 \pm 0.31 \pm
0.40)$~MeV/$c^2$ is obtained, in good agreement with the inclusive measurement. The
exclusive reconstruction allows to determine the $J^P$ quantum number: fits to the
helicity distribution in the data favor the quantum number $J=1$ while $J=2$ is
disfavored. More statistics is needed to conclude definitely and determine the parity
$P$.

\subsection{$D_{s0}^*(2317)$ and $D_{s1}(2460)$ mesons}

In 2003, two new resonances were discovered by the \bbrExp\ and CLEO experiments: the
$D_{s0}^*(2317)$ and $D_{s1}(2460)$ mesons\cite{ref:firstDsJ}. These two resonances are
very narrow, and have masses well below what was predicted by the potential model (the
$D_{s0}^*(2317)$ and $D_{s1}(2460)$ mesons were predicted to lie above the $DK$ and
$D^*K$ threshold, respectively, but have been observed below these thresholds). These
states are very well known experimentally: masses are measured with an error below
2~MeV/$c^2$, 95\% confidence level upper limits on widths are about 4~MeV~; $J^P$ quantum
numbers ($0^+$ and $1^+$ for $D_{s0}^*(2317)$ and $D_{s1}(2460)$ respectively), decay
modes and branching fractions are also well measured. Despite a good knowledge of these
states, their theoretical interpretation is still unclear. One obvious possibility is to
identify these two resonances with the $0^+$ and $1^+$ $c\bar{s}$ states, although it is
difficult to fit these resonances within the potential model. Other interpretations have
been proposed: four quark states, $DK$ molecules or $D\pi$
atoms\cite{ref:firstDsJModels}.

\subsection{$D_{sJ}^*(2860)$ meson}

The $D_{sJ}^*(2860)$ resonance was discovered by \bbrExp\ in 2006\cite{ref:DsJAntimo},
looking in the $c\bar{c}$ continuum: $e^+e^- \to D^0 K^+ X$ and $e^+e^- \to D^+ K^0_s X$,
where we consider the decays $D^0 \to K^- \pi^+, K^- \pi^+ \pi^0$ and $D^+ \to K^- \pi^+
\pi^+$ and where $X$ could be anything. A clear peak is observed in the $DK$ invariant
mass for the sum of these 3 modes, with a mass of $(2856.6 \pm 1.5 \pm 5.0)$~MeV/$c^2$
and a width of $(47 \pm 7 \pm 10)$~MeV. This signal is seen with a statistical
significance above $8\sigma$. Although this resonance is observed with a high
significance, no other experiments have yet confirmed this state. Given that this
resonance decays to two pseudoscalars, the $J^P$ quantum number should be $0^+$, $1^-$,
$2^+$, etc. Different interpretations have been proposed, inside the $c\bar{s}$ scheme:
this state could be a radial excitation of the $D_{s0}^*(2317)$, but other possibilities
are not ruled out\cite{ref:DsJAntimoModels}.

\subsection{$D_{sJ}(2700)$ meson}

In the same analysis, \bbrExp\ reported a broad enhancement, named $X(2690)$, at a mass
of $(2688 \pm 4 \pm 3)$~MeV/$c^2$ and a width of $(112 \pm 7 \pm 36)$~MeV. A new state,
the $D_{sJ}(2700)$, was reported independently by Belle at a similar mass, with a $J^P$
quantum number equal to $1^-$, looking at $B^+ \to \bar{D}^0 D^0 K^+$
events\cite{ref:DsJ2700}. Since the $X(2690)$ and $D_{sJ}(2700)$ mesons have the same
decay modes and that the mass and width are consistent with each other, it is reasonable
to think that they are indeed the same state.

\bbrExp\ performed an exclusive analysis\cite{ref:DsJ2700}, looking at events where $B$
decays to $\bar{D}^{(*)} D^{(*)} K$. Thanks to the many final states studied, this
analysis has the advantage to be able to look at four $D^0K^+$ invariant mass
distributions as well as four $D^+K^0_s$ invariant mass distributions. Adding these final
states together, a clear resonant enhancement is seen around a mass of 2700~MeV/$c^2$.
Furthermore, adding the four $D^{*0}K^+$ and four $D^{*+}K^0_s$ invariant mass
distributions together, a similar enhancement is observed around a mass of
2700~MeV/$c^2$. No precise measurement was given by this preliminary analysis yet.

The potential model predicts the $2^3S_1$ $c\bar{s}$ state at a mass of 2720~MeV/$c^2$.
Also, from chiral symmetry considerations, a $1^+-1^-$ doublet of states has been
predicted. If the $1^+$ state is identified as the $D_{s1}(2536)$, the mass predicted for
the $1^-$ state is $(2721 \pm 10)$~MeV/$c^2$\cite{ref:DsJ2700Models}.

\section{Charmed baryons}

The production and decay of singly-charmed baryons are largely unexplored and provide an
interesting environment to study the dynamics of quark-gluon interactions. All nine
ground states with $J^P=1/2^+$ and all six ground states with $J^P=3/2^+$ are now
observed (the last missing state, $\Omega_c^{*0}$, was discovered by \bbrExp\ in 2006).
Several orbitally excited states have already been seen as well. The spin and parity of
each of these excited single-charm baryons are assigned based on a comparison of the
measured masses and natural widths with predictions of theoretical models.

\subsection{$\Omega_c^0$ and $\Omega_c^{*0}$ baryons}

The $\Omega_c^0$ baryon, a $css$ state with $J^P=1/2^+$, was observed by
\bbrExp\cite{ref:omega_c} decaying to four hadronic modes: $\Omega_c^0 \to \Omega^-
\pi^+$ (with a significance of $18\sigma$), $\Omega_c^0 \to \Omega^- \pi^+ \pi^0$
(5.1$\sigma$), $\Omega_c^0 \to \Omega^- \pi^+ \pi^+ \pi^-$ ($4.2\sigma$) and $\Omega_c^0
\to \Xi^- K^- \pi^+ \pi^+$ ($4.3\sigma$). The ratios of branching fractions were
measured, significantly improving upon the previous values. The $p^*$ spectrum of
$\Omega_c^0$ was also measured in order to study the production rates in both $c\bar{c}$
and $B\bar{B}$ events, using only the $\Omega^- \pi^+$ final state. This analysis find
comparable production rates of $\Omega_c^0$ baryons from the continuum and from $B$ meson
decays. This is the first observation of this baryon in $B$ decays.

Recently, \bbrExp\ discovered\cite{ref:omega_cstar} the $css$ ground state with
$J^P=3/2^+$, the $\Omega_c^{*0}$ baryon. This baryon was observed with a significance of
$5.2\sigma$ in the decay to $\Omega_c^0 \gamma$, combining the four decay modes of the
$\Omega_c^0$ cited in the previous paragraph. The difference of mass between
$\Omega_c^{*0}$ and $\Omega_c^0$ was measured to be $(70.8 \pm 1.0 \pm 1.1)$~MeV/$c^2$. A
non-relativistic QCD effective field theory calculation predicts the mass difference to
be in the range 50-73~MeV/$c^2$, while a lattice calculation gives a mass difference
equal to $(94 \pm 10)$~MeV/$c^2$~\cite{ref:omega_theory}.

The ratio of inclusive production cross-section was determined to be
\begin{equation}
\frac{\sigma(e^+e^- \to \Omega_c^{*0} X, x_p(\Omega_c^{*0})>0.5)}{\sigma(e^+e^- \to
\Omega_c^{0} X, x_p(\Omega_c^{0})>0.5)} = 1.01 \pm 0.23 \pm 0.11,
\end{equation}
where the scaled momentum $x_p = p^*/p^*_{\mathrm{max}}$ of the $\Omega_c^{*0}$
($\Omega_c^{0}$) is required to be greater than 0.5 in the numerator (denominator)
cross-section.

\subsection{$\Lambda_c(2940)^+$ baryon}

A search for charmed baryons decaying to $D^0 p$ was performed\cite{ref:lambda_c} and
revealed two states: the $\Lambda_c(2880)^+$ baryon and a previously unobserved state,
the $\Lambda_c(2940)^+$. This is the first observation of charmed baryons decaying to a
$D$ meson and a light baryon.

The $\Lambda_c(2880)^+$ baryon was previously observed\cite{ref:lambda_cleo} by the CLEO
experiment in the $\Lambda_c \pi^+ \pi^-$ decay mode, with a mass of $(2880.9 \pm
2.3)$~MeV/$c^2$ and a width less than 8~MeV at 90\% confidence level. Using \bbrExp\ data
in the $D^0 p$ channel leads to much more precise values, in particular with the first
measurement of the width of the $\Lambda_c(2880)^+$: $m = (2881.9 \pm 0.1 \pm
0.5)$~MeV/$c^2$ and $\Gamma = (5.8 \pm 1.5 \pm 1.1)$~MeV. The existence of the decay $D^0
p$~ rules out various interpretations of this baryon\cite{ref:lambda_theory}.

The new baryon $\Lambda_c(2940)^+$ is observed with a significance above $7\sigma$, with
a mass of $(2939.8 \pm 1.3 \pm 1.0)$~MeV/$c^2$ and an intrinsic width of $(17.5 \pm 5.2
\pm 5.9)$~MeV. This new state could be interpreted as a $udc$ baryon. To determine if
this new state belongs to an isotriplet (analogous to $\Sigma_c^0$ and $\Sigma_c^{++}$),
a search for doubly-charged partner was performed, looking at the decay mode $D^+p$. No
signal corresponding to either the $\Lambda_c(2880)^+$ or $\Lambda_c(2940)^+$ baryon was
observed, which shows that both states are isoscalar.

\subsection{$\Xi_c(2980)^+$, $\Xi_c(3077)^{+/0}$, $\Xi_c(3055)^+$ and $\Xi_c(3123)^+$ baryons}

The \bbrExp\ experiment searched\cite{ref:xi_c} for the excited charm-strange baryons
$\Xi_c(2980)^+$ and $\Xi_c(3077)^{+/0}$, discovered previously by the Belle
collaboration\cite{ref:xi_belle}. \bbrExp\ confirms the states $\Xi_c(2980)^+$,
$\Xi_c(3077)^{+/0}$, looking at the $\Lambda_c^+ K^- \pi^+$ and $\Lambda_c^+ K^0_S \pi^-$
final states, with $\Lambda_c^+ \to p K^- \pi^+,~ p K^0_S,~ p K^0_S \pi^-\pi^+,~ \Lambda
\pi^+,~ \Lambda \pi^+ \pi^- \pi^+$. The $\Xi_c(2980)^+$ baryon is observed at a mass of
$(2969.3 \pm 2.2 \pm 1.7)$~MeV/$c^2$ with a width of $(27 \pm 8 \pm 2)$~MeV. The
$\Xi_c(3077)^{+}$ is seen at a mass of $(3077.0 \pm 0.4 \pm 0.2)$~MeV/$c^2$ with an
intrinsic width of $(5.5 \pm 1.3 \pm 0.6)$~MeV. \bbrExp\ confirmed also the
$\Xi_c(3077)^{0}$ state at a mass of $(3079.3 \pm 1.1 \pm 0.2)$~MeV/$c^2$ with a width of
$(5.9 \pm 2.3 \pm 1.5)$~MeV. These results are in good agreement with the values given by
Belle, except for the $\Xi_c(2980)^+$ baryon where the difference could be explained by
the fact that \bbrExp\ incorporates phase space effect near the threshold and takes into
account decays to $\Sigma_c(2455)^{++} K^-$.

In addition to these confirmations, \bbrExp\ discovered a new baryon, the
$\Xi_c(3055)^+$, and found an evidence for the $\Xi_c(3123)^+$ baryon. The
$\Xi_c(3055)^+$ and $\Xi_c(3123)^+$ signals are observed only in $\Sigma_c(2455)^+K^-$
and $\Sigma_c(2520)^+ K^-$ intermediate-resonant decays, respectively. The
$\Xi_c(3055)^+$ is seen with a $6.4\sigma$ significance, with a mass of $(3054.2 \pm 1.2
\pm 0.5)$~MeV/$c^2$ and a width of $(17 \pm 6 \pm 11)$~MeV, while the $\Xi_c(3123)^+$ is
observed with a $3.6\sigma$ significance at a mass of $(3122.9 \pm 1.3 \pm
0.3)$~MeV/$c^2$ and with a width of $(4.4 \pm 3.4 \pm 1.7)$~MeV.

These baryons have same or similar decay channels as the double-charm baryon, but are
identified as single-charm based on the measured masses, natural widths and charges of
the members of the isospin doublet. In the current state of knowledge, it is difficult to
assign a spin-parity to these baryons. More theoretical and experimental work is needed
to clarify the properties of these states.

\section{Conclusion}

Although no new resonances were discovered in many years, \bbrExp\ gave an impressive
list of new results since 1999. In the $c\bar{s}$ sector, the $D_{s0}^*(2317)$ and
$D_{s1}(2460)$ mesons are now very well known experimentally, but no definite
interpretation was given theoretically. The $D_{sJ}^*(2860)$ and $D_{sJ}(2700)$ mesons
were discovered recently and need more experimental inputs. Many new charmed baryon
states were observed by \bbrExp. Thanks to the discovery of the $\Omega_c^{*0}$, all
ground states are now established. A lot of excited states have been observed, and
probably many of them have yet to be discovered. The production rate, the decay channels
and the measured properties (masses, widths) of these excited states will help to
understand the internal quark dynamics.

A lot of analyzes are still in progress with the current data set in \bbrExp: more decay
modes for the resonances presented here are being investigated. \bbrExp\ is taking data
until the end of 2008, which is the promise of more surprises to arise.

\section{Acknowledgements}

The author is very grateful to the organizers of the HADRON 2007 conference for their
support and all efforts in making this venue successful.

We are also grateful for the excellent luminosity and machine conditions provided by our
PEP-II colleagues, and for the substantial dedicated effort from the computing
organizations that support \bbrExp.


\begin{thebibliography}{99}
%
\bibitem{ref:babar} B. Aubert {\it et al.} (\bbrExp\ Collaboration), Nucl. Instrum. Meth. {\bf A 479} 1
(2002).
\bibitem{ref:potential} S. Godfrey and N. Isgur, Phys. Rev. {\bf D 32} 189 (1985).
\bibitem{ref:Ds1} B. Aubert {\it et al.} (\bbrExp\ Collaboration), arXiv:hep-ex/0607084 (2006).
\bibitem{ref:DDK} B. Aubert {\it et al.} (\bbrExp\ Collaboration), arXiv:0708.1565 (2007).
\bibitem{ref:firstDsJ} S.K. Choi {\it et al.} (Belle Collaboration), Phys. Rev. Lett. {\bf 91}
262001(2003)~; B. Aubert {\it et al.} (\bbrExp\ Collaboration), Phys. Rev. {\bf D74}
032007 (2006).
\bibitem{ref:firstDsJModels} H-Y Cheng and W-S Hou, Phys. Lett. {\bf B566} 193 (2003)~;
T. Barnes, F. E. Close, and H. J. Lipkin, Phys. Rev. {\bf D68} 054006 (2003)~; A.
Szczepaniak, Phys. Lett. {\bf B567} 23 (2003).
\bibitem{ref:DsJAntimo} B. Aubert {\it et al.} (\bbrExp\ Collaboration), Phys. Rev. Lett. {\bf97} 222001 (2006).
\bibitem{ref:DsJAntimoModels} E. van Beveren, and G. Rupp, Phys. Rev. Lett. {\bf 97}
202001 (2006)~; P. Colangelo, F. De Fazio, and S. Nicotri, Phys. Lett. {\bf B642} 48
(2006).
\bibitem{ref:DsJ2700} J. Brodzicka {\it et al.} (Belle Collaboration), arXiv:0707.3491 (2007)~;
B. Aubert {\it et al.} (\bbrExp\ Collaboration), preliminary (2007).
\bibitem{ref:DsJ2700Models} M. A. Nowak, M. Rho, and I. Zahed, Phys. Polon. {\bf B 35}, 2377 (2004).
\bibitem{ref:omega_c} B. Aubert {\it et al.} (\bbrExp\ Collaboration), Phys. Rev. Lett. {\bf 99} 062001 (2007).
\bibitem{ref:omega_cstar} B. Aubert {\it et al.} (\bbrExp\ Collaboration), Phys. Rev. Lett. {\bf 97} 232001 (2006).
\bibitem{ref:omega_theory} N. Mathur {\it et al.}, Phys. Rev. Lett. {\bf B592}, 1
(2004)~; R. M. Woloshyn, Nucl. Phys. Proc. Suppl. {\bf 93}, 38 (2001).
\bibitem{ref:lambda_c} B. Aubert {\it et al.} (\bbrExp\ Collaboration), Phys. Rev. Lett. {\bf 98}, 012001 (2007).
\bibitem{ref:lambda_cleo} M. Artuso {\it et al.} (CLEO), Phys. Rev. Lett. {\bf 86}, 4479
(2001).
\bibitem{ref:lambda_theory} A. E. Blechman, A. F. Falk, D. Pirjol, and J. M. Yelton,
Phys. Rev. {\bf D67}, 074033 (2003).
\bibitem{ref:xi_c} B. Aubert {\it et al.} (\bbrExp\ Collaboration), arXiv:hep-ex/0607042 (2006)~;
 B. Aubert {\it et al.} (\bbrExp\ Collaboration), preliminary (2007).
 \bibitem{ref:xi_belle} R. Chistov {\it et al.} (Belle Collaboration),
 arXiv:hep-ex/0606051 (2006).

\end{thebibliography}
\end{document}